\documentclass{article}
\usepackage{spconf,amsmath,graphicx}

\usepackage{hyperref}
\usepackage{booktabs}
\usepackage{multirow}
\usepackage{caption}
\usepackage{subcaption}
\usepackage{enumitem}
\usepackage{pbox}
\usepackage{tikz}

\title{Emotion-Aligned Contrastive Learning Between Images and Music}

\name{Shanti Stewart$^1$ \qquad Kleanthis Avramidis$^{1, {\star}}$ \qquad Tiantian Feng$^{1, {\star}}$ \qquad Shrikanth Narayanan$^1$}
\address{$^1$ Signal Analysis and Interpretation Lab, University of Southern California, USA}

\begin{document}
\ninept 

\maketitle

\def\thefootnote{*}\footnotetext{These authors contributed equally to this work.}\def\thefootnote{\arabic{footnote}}
\def\thefootnote{}\footnotetext{This work was supported by the USC Center for Computational Media Intelligence and its sponsors.}\def\thefootnote{\arabic{footnote}}

\begin{abstract}
Traditional music search engines rely on retrieval methods that match natural language queries with music metadata. There have been increasing efforts to expand retrieval methods to consider the audio characteristics of music itself, using queries of various modalities including text, video, and speech. While most approaches aim to match general music semantics to the input queries, only a few focus on affective qualities. In this work, we address the task of retrieving emotionally-relevant music from image queries by learning an affective alignment between images and music audio. Our approach focuses on learning an emotion-aligned joint embedding space between images and music. This embedding space is learned via emotion-supervised contrastive learning, using an adapted cross-modal version of the SupCon loss. We evaluate the joint embeddings through cross-modal retrieval tasks (image-to-music and music-to-image) based on emotion labels. Furthermore, we investigate the generalizability of the learned music embeddings via automatic music tagging. Our experiments show that the proposed approach successfully aligns images and music, and that the learned embedding space is effective for cross-modal retrieval applications.
\end{abstract}


\section{Introduction}
\label{sec:Introduction}

Modern large-scale music search engines primarily retrieve music by matching natural language queries with music metadata---such as the artist's name, album title, or song title. While some of these retrieval systems allow querying by genre or mood, they often fall short in supporting high-granularity queries. Users specify their queries in a pre-defined set of descriptors, such as ``jazz" (genre) and ``happy" (mood), instead of detailed musical descriptions (e.g., ``a happy upbeat Latin jazz song with saxophone and bass"). In addition, existing music retrieval systems typically focus on metadata and do not consider the auditory characteristics of the music.

There have been increasing efforts to address this problem.  Won et al. \cite{Won-2021-tag_based_music_retrieval} present a method to retrieve music audio from single-word (tag) queries. Manco et al. \cite{Manco-2022-MusCALL} instead propose a framework for cross-modal text-to-music retrieval from free-form sentence queries. Doh et al. \cite{Doh-2023-universal_text_to_music_retrieval} combine both tag-based and sentence-based music retrieval methods into a unified framework. In addition, there have been a number of works on video-to-music retrieval \cite{Li-2019-query_by_video, Suris-2022-artistic_correspondence, Cheng-2023-SSVMR}.

These newer cross-modal music retrieval frameworks operate on general audio semantics and typically use paired multimodal datasets \cite{Manco-2022-MusCALL, Li-2019-query_by_video, Suris-2022-artistic_correspondence} or some form of weak language supervision \cite{Won-2021-tag_based_music_retrieval, Doh-2023-universal_text_to_music_retrieval}. While the paired datasets can sometimes be organically created when two modalities co-occur naturally (e.g., video and music in music videos), the pairings are often generated by human annotators. Such semantic pairings can be subjective, and manual annotation is costly in time and effort.

An alternative to retrieving music based on general semantics is through cross-modal class supervision. Finding semantic classes that are compatible across multiple modalities is challenging; classes used in one modality (e.g., image object classes) may not have equivalent meanings in other modalities.
Emotions, however, have equivalent meanings across multiple modalities: images, language, speech, and music. On this idea, two different works present methods for emotion-supervised cross-modal music retrieval. Won et al. \cite{Won-2021-emotion_embedding_spaces} propose a framework for text-to-music retrieval based on emotions, and Doh et al. \cite{Doh-2023-speech_to_music} extend this method for speech-to-music retrieval.

Building upon this body of work, we address the task of emotion-supervised music retrieval from image queries. To the best of our knowledge, this problem has not been previously addressed in the literature. Retrieving emotionally-relevant music from images introduces several benefits. Using non-language queries is sometimes more intuitive,
and automatically matching emotionally-similar images and music can encourage the creation of more compelling multimedia content.

To this end, we propose \textit{Emo-CLIM}: a framework for \underline{Emo}tion-Aligned \underline{C}ontrastive \underline{L}earning Between \underline{I}mages and \underline{M}usic.\footnote{Code is available at \href{https://github.com/shantistewart/Emo-CLIM}{https://github.com/shantistewart/Emo-CLIM}} Our approach learns an emotion-aligned joint embedding space between images and music, in which embeddings of emotionally-similar images and music are close together. We then directly leverage these joint embeddings for emotion-supervised cross-modal retrieval. In contrast to prior work \cite{Won-2021-emotion_embedding_spaces, Doh-2023-speech_to_music} which use triplet loss functions, we use a supervised contrastive loss—which has the
benefit of comparing across all items in a training batch. Furthermore, our loss is modality-symmetric, unlike \cite{Won-2021-emotion_embedding_spaces, Doh-2023-speech_to_music}, allowing the embedding space to be used for both image-to-music and music-to-image retrieval. Our key contributions can be summarized as follows:

\begin{itemize}[leftmargin=*]
\item  To the best of our knowledge, Emo-CLIM is the first framework that learns an affective alignment between images and music audio. This framework is distinct from existing literature that aligns music with other modalities.

\item Unlike prior work that uses triplet losses, Emo-CLIM uses an emotion-supervised contrastive loss, demonstrating promising results in cross-modal retrieval as well as automatic music tagging.
\end{itemize}

\section{Related Work}
\label{sec:Related Work}




Many works have successfully applied contrastive learning to multimodal problems. CLIP \cite{Radford-2021-CLIP} used contrastive learning between images and text to learn effective image representations, and AudioCLIP \cite{Guzhov-2022-AudioCLIP} and Wav2CLIP \cite{Wu-2022-Wav2CLIP} extended CLIP to handle audio. Several other studies have explored contrastive learning to align language and audio \cite{Elizalde-2022-CLAP_original, Wu-2023-CLAP}.
There have also been a number of works using multimodal contrastive learning in the music domain. MusCALL \cite{Manco-2022-MusCALL} and MuLan \cite{Huang-2022-MuLan} proposed contrastive learning approaches between language and music audio, and several other works  \cite{Suris-2022-artistic_correspondence, Avramidis-2023-VCMR} explored similar approaches for videos and music.


A common application for multimodal embedding spaces is cross-modal retrieval. Several works \cite{Manco-2022-MusCALL, Huang-2022-MuLan, Doh-2023-universal_text_to_music_retrieval} learn joint embedding spaces between language and music audio, which are used for text-to-music retrieval. Methods for music retrieval from video queries have also been proposed \cite{Li-2019-query_by_video, Suris-2022-artistic_correspondence, Cheng-2023-SSVMR}. Although there are numerous papers on cross-modal music retrieval, music retrieval based on emotions---the focus of our work---is under-explored. Among studies that address this topic, Won et al.~\cite{Won-2021-emotion_embedding_spaces} implement text-to-music retrieval, and Doh et al.~\cite{Doh-2023-speech_to_music} implement speech-to-music retrieval.

\section{Emo-CLIM Framework}
\label{sec:Emo-CLIM Framework}

\begin{figure}[t]
    \begin{center}
       \includegraphics[width=0.9\linewidth]{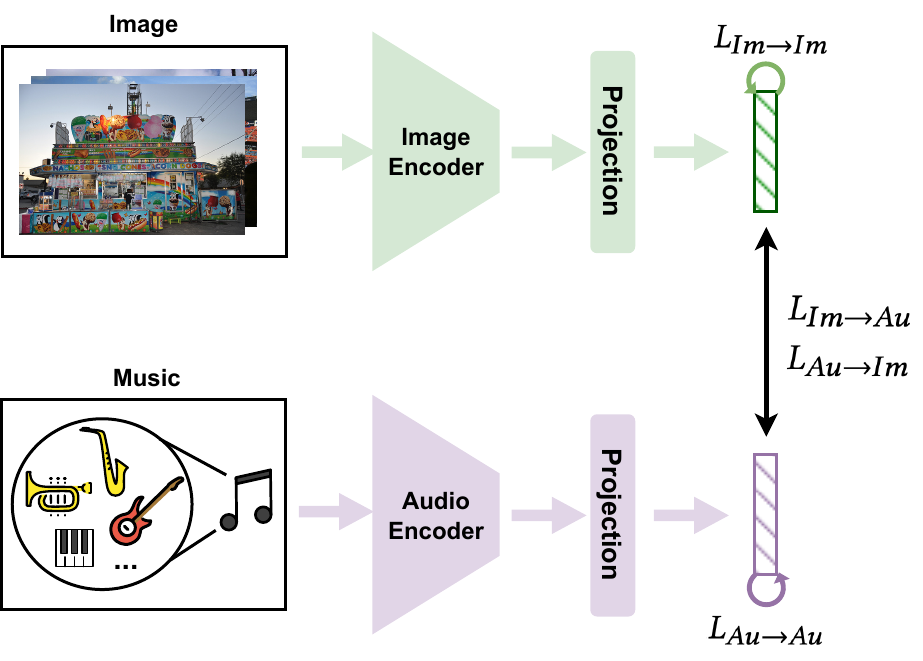}
    \end{center}
    \vspace{-3mm}
    \caption{Overview of the Emo-CLIM framework. A dual-branch architecture separately encodes images and music, then projects the encoded features to an emotion-aligned joint embedding space. Two cross-modal (image-to-audio and audio-to-image) and two intra-modal (image-to-image and audio-to-audio) contrastive losses operate on the joint embeddings.}
    \vspace{-2mm}
    \label{fig:framework}
\end{figure}

As shown in Figure \ref{fig:framework}, the Emo-CLIM framework consists of three main components: feature extraction, modality alignment, and emotion-supervised contrastive learning. Given an image $ x^{(Im)} $ and an audio (music) clip $ x^{(Au)} $, Emo-CLIM computes an image embedding $ z^{(Im)} $ and an audio embedding $ z^{(Au)} $ as follows:
\begin{equation}
\label{eq:framework}
\footnotesize
z^{(Im)} = h_{Im} \left(f_{Im} \left(x^{(Im)}\right)\right); z^{(Au)} = h_{Au} \left(f_{Au} \left(x^{(Au)}\right)\right)
\end{equation}
where $ f_{Im}(\cdot) $ and $ f_{Au}(\cdot) $ are image and audio encoders, and $ h_{Im}(\cdot) $ and $ h_{Au}(\cdot) $ are projection networks for the image and audio modalities, respectively. The encoder networks extract modality-specific features, and the projection networks map these features to a joint embedding space. We use supervised contrastive learning to align emotionally-paired images and audio clips in this embedding space.

\subsection{Feature Extraction}
\label{subsec:Feature Extraction}


For the image encoder $ f_{Im}(\cdot) $, we use the vision transformer component of the CLIP model \cite{Radford-2021-CLIP}. We obtain the pre-trained model from OpenAI's official GitHub repository.\footnote{\href{https://github.com/openai/CLIP}{https://github.com/openai/CLIP}}
During training, we keep the CLIP model frozen, since CLIP embeddings have been shown to be effective without fine-tuning \cite{Radford-2021-CLIP}, and our datasets are too small to fine-tune a model of this size.


For the audio encoder $ f_{Au}(\cdot) $, we use music-specific and general audio representation models.
For the music-specific models, we use two different architectures that are commonly used in the music information retrieval domain: Short-Chunk CNN \cite{Won-2020-music_tagging_evaluation} and Harmonic CNN \cite{Won-2020-HarmonicCNN}. Both are CNN-based architectures and take in mel-spectrogram inputs.
Short-Chunk CNN operates on approximately 3.7-second input audio clips, while Harmonic CNN operates on 5.0-second audio clips \cite{Won-2020-music_tagging_evaluation}. We utilize pre-trained models---trained on automatic music tagging using the Million Song Dataset \cite{Mahieux-2011-MSD}---and obtain model weights from an open-source repository.\footnote{\href{https://github.com/minzwon/sota-music-tagging-models}{https://github.com/minzwon/sota-music-tagging-models}}

For the general audio model, we use the audio component of the CLAP model \cite{Wu-2023-CLAP}. CLAP is a transformer-based model that operates on an audio input of 10.0 seconds. We download the pre-trained model weights from an open-source GitHub repository\footnote{\href{https://github.com/LAION-AI/CLAP}{https://github.com/LAION-AI/CLAP}}, and select the checkpoint that was trained without AudioSet data to ensure a fair evaluation. We keep the CLAP model frozen during training.

\subsection{Modality Alignment}
\label{subsec:Modality Alignment}

To map the image and audio features to the joint embedding space, we use two separate projection networks---one for each modality. Each network is a small multi-layer perceptron (MLP), consisting of a linear layer, batch normalization layer, ReLU activation, dropout layer, and a second linear layer that yields 128-dimensional embeddings (which are $ L_2 $-normalized).

\begin{table*}[t]
    \caption{Cross-modal and intra-modal retrieval performance on the DeepEmotion \cite{You-2016-DeepEmotion} image dataset and the AudioSet music mood subset \cite{Gemmeke-2017-AudioSet}, for five different audio encoder variants. We report Precision@5 (P@5) and Mean Reciprocal Rank (MRR) evaluation metrics. A retrieved item is considered correct if it has the same emotion label as the query.}
    \label{table:retrieval_performance}
    \footnotesize
    \centering
    
    \begin{tabular}{l | l l | l l | l l | l l}
        \toprule
        
        \multirow{2}{*}{Audio Encoder Model} &
        \multicolumn{2}{c |}{Image $ \rightarrow $ Music} &
        \multicolumn{2}{c |}{Music $ \rightarrow $ Image} &
        \multicolumn{2}{c |}{Image $ \rightarrow $ Image} &
        \multicolumn{2}{c}{Music $ \rightarrow $ Music}
        \\
        
        & P@5 & MRR & P@5 & MRR &
        P@5 & MRR & P@5 & MRR
        \\
        \midrule
        
        Short-Chunk CNN (Frozen) & 64.23\% & 71.88\% & 61.43\% & 70.78\% & \textbf{72.54\%} & \textbf{81.07\%} & 55.34\% & 68.90\%
        \\
        
        Short-Chunk CNN (Unfrozen) & 63.95\% & 75.50\% & 63.46\% & 69.87\% & 69.25\% & 78.72\% & 55.15\% & 67.43\%
        \\
        
        Harmonic CNN (Frozen) & 65.94\% & \textbf{78.59\%} & 64.34\% & 72.23\% & 70.48\% & 79.54\% & 55.16\% & 68.46\%
        \\
        
        Harmonic CNN (Unfrozen) & 63.18\% & 74.08\% & \textbf{67.58\%} & \textbf{74.0\%} & 68.64\% & 78.20\% & 57.83\% & 68.46\%
        \\
        
        CLAP (Frozen) & \textbf{68.15\%} & 76.65\% & 67.32\% & 73.95\% & 70.71\% & 79.27\% & \textbf{60.80\%} & \textbf{72.19\%}
        \\
        
        \bottomrule
    
    \end{tabular}

\end{table*}

\subsection{Emotion-Supervised Contrastive Learning}
\label{subsec:Emotion-Supervised Contrastive Learning}

To learn the emotion-aligned multimodal embedding space, we use supervised contrastive learning on the joint embeddings, supervised by emotion labels. To this end, we adapt the SupCon loss \cite{Khosla-2020-SupCon} to our multimodal setting, as follows.

Given a batch of $ N $ images with their emotion labels \\ $ \{ ( x_i^{(Im)}, \ y_i^{(Im)} ) \}_{i=1}^{N} $ and $ N $ music audio clips with their emotion labels $ \{ ( x_j^{(Au)}, \ y_j^{(Au)} ) \}_{j=1}^{N} $, we compute 4 different supervised contrastive losses, as detailed in the following subsections. For the remainder of this paper, we adopt the following notations: $ y_i^{(M)} $ = emotion label of sample $ i $ of modality $ M $, $ z_i^{(M)} $ = embedding of sample $ i $ of modality $ M $, $ I = \{ 1,...,N \} $ = all indices in a batch, and $ \tau $ = the temperature hyperparameter.

\textbf{Cross-Modal Contrastive Losses}:
To align the image and audio modalities, we use a cross-modal version of the SupCon loss. Given $ N $ samples $ \{ (x_i^{(M_1)}, \ y_i^{(M_1)}) \}_{i=1}^{N} $ from modality $ M_1 $ and $ N $ samples $ \{ (x_p^{(M_2)}, \ y_p^{(M_2)}) \}_{p=1}^{N} $ from modality $ M_2 $, our cross-modal $ M_1 \rightarrow M_2 $ SupCon loss is:
\begin{equation}
\label{eq:cross_modal_supcon}
\begin{split}
L_{M_1 \rightarrow M_2} = - \frac{1}{N} \sum_{i=1}^{N} \frac{1}{|P^{(M_1 \rightarrow M_2)}(i)|}
\\
\sum_{p \in P^{(M_1 \rightarrow M_2)}(i)} log \ \frac{exp(z_i^{(M_1)} \cdot z_p^{(M_2)} / \tau)} {\sum_{k \in I} exp(z_i^{(M_1)} \cdot z_k^{(M_2)} / \tau) }
\end{split}
\end{equation}

$ P^{(M_1 \rightarrow M_2)}(i) $ is the set of indices of positive samples $ x_p^{(M_2)} $ for anchor sample $ x_i^{(M_1)} $, and is defined as:
\begin{equation}
\label{eq:cross_modal_supcon_index_sets}
\footnotesize
P^{(M_1 \rightarrow M_2)}(i) = \{ p \in I \ | \ y_i^{(M_1)} = y_p^{(M_2)} \}
\end{equation}

These cross-modal SupCon losses ``pull together" cross-modal embeddings with the same emotion label and ``push apart" cross-modal embeddings with different emotion labels.

\textbf{Intra-Modal Contrastive Losses}:
To learn a more robust joint embedding space as well as regularize the cross-modal objectives, we include intra-modal SupCon loss terms in our full objective. The intra-modal SupCon losses are defined as in \autoref{eq:cross_modal_supcon} with $ M_1 = M_2 $ . These intra-modal SupCon losses ``pull together" same-modality embeddings with the same emotion label and ``push apart" same-modality embeddings with different emotion labels.

\textbf{Total Contrastive Loss}:
The total combined loss is a weighted average of 2 cross-modal and 2 intra-modal losses:
\begin{equation}
\label{eq:total_supcon}
\footnotesize
L_{\text{total}} = \lambda_1 L_{Im \rightarrow Au} + \lambda_2 L_{Au \rightarrow Im}
+ \lambda_3 L_{Im \rightarrow Im} + \lambda_4 L_{Au \rightarrow Au}
\end{equation}

$ L_{\text{total}} $ is modality-symmetric, which ensures the joint embedding space does not favor one modality over the other.
Thus, the adapted supervised contrastive objective enables us to learn a joint embedding space between images and music audio, aligned both in an intra-modal and cross-modal manner.

\section{Experiments and Results}
\label{sec:Experiments and Results}

\subsection{Datasets}
\label{subsec:Datasets}


We use the DeepEmotion image dataset \cite{You-2016-DeepEmotion}, which consists of 21,829
annotated images collected from Flickr and Instagram. Each image is assigned a single emotion label among 8 labels: \emph{amusement}, \emph{awe}, \emph{contentment}, \emph{excitement}, \emph{anger}, \emph{disgust}, \emph{fear}, and \emph{sadness}. To create training/validation/test subsets, we use a random 80-10-10\% split, stratified with respect to the labels.

For the music dataset, we use the AudioSet music mood subset \cite{Gemmeke-2017-AudioSet}, which consists of 13,713
10.0-second music (audio) clips gathered from YouTube. Each music clip is assigned a single emotion label among 7 labels: \emph{exciting}, \emph{funny}, \emph{happy}, \emph{tender}, \emph{angry}, \emph{sad}, and \emph{scary}. To create training/validation/test subsets, we likewise use a random 80-10-10\% split, stratified with respect to the labels.


The emotion label taxonomies of the image and music datasets are different. To address this issue, we define a manual mapping between these labels\footnote{Mapping available at \href{https://github.com/shantistewart/Emo-CLIM}{https://github.com/shantistewart/Emo-CLIM}.}, many of which differ only in wording (e.g., \emph{excitement} and \emph{exciting}). However, \emph{awe} and \emph{disgust} images and \emph{tender} music do not have clear equivalents. Hence, we completely remove all images/audio clips with these three emotion labels in order to avoid ambiguous or illogical mappings.


\subsection{Implementation Details}
\label{subsec:Implementation Details}


Since we use CLIP to encode images, we apply the corresponding image pre-processing transforms\footnote{Details can be found at \href{https://github.com/openai/CLIP}{https://github.com/openai/CLIP}.}, which include resizing and cropping to a size of $ 224 \times 224 $ and normalization. We use random cropping during training and center cropping during evaluation.


We use raw audio at a sampling rate of 16 kHz.\footnote{When using CLAP, we up-sample to the expected 48 kHz sampling rate.} Since AudioSet contains 10.0-second audio clips and the music-specific audio encoder models operate on shorter inputs, we randomly crop audio segments during training. During evaluation, we use a sliding window with an overlap ratio of 75\% to divide each 10.0-second audio clip into multiple segments,
then compute the average embedding over all segments.
When using CLAP, we do not use these methods, since CLAP's input length matches AudioSet.


We set the dimension of the joint embedding space to 128.
For the contrastive losses, we use a temperature of 0.07 and equal $ \lambda $ values ($ \lambda_1 = \lambda_2 = \lambda_3 = \lambda_4 = 0.25 $).
For all experiments, we use the AdamW \cite{Loshchilov-2019-AdamW} optimizer with a batch size of 64 and a learning rate of 0.0001. We train all models for 15 epochs and keep the checkpoint with the lowest validation loss.

\subsection{Cross-Modal Retrieval}
\label{subsec:Cross-Modal Retrieval}

Following other works \cite{Manco-2022-MusCALL, Huang-2022-MuLan, Won-2021-emotion_embedding_spaces, Doh-2023-speech_to_music}, we evaluate the learned joint embedding space via cross-modal retrieval. Given a query item of one modality, we retrieve the most similar item of the other modality using a simple nearest-neighbor search. We use the held-out test subsets of the image and music datasets for all retrieval evaluations.

\textbf{Experimental Setup}:
Cross-modal retrieval evaluation is implemented as a ranking problem. Given a query item of one modality, we rank all items of the other modality by the cosine similarity between the query and candidate item embeddings. We report Precision@5 (P@5) and Mean Reciprocal Rank (MRR) scores.
A retrieved item is considered correct if it has the same emotion label as the query. In line with \cite{Won-2021-emotion_embedding_spaces}, we macro-average retrieval metrics across emotion classes in order to avoid potential bias caused by the imbalanced emotion class distributions.

\textbf{Results}:
\autoref{table:retrieval_performance} presents cross-modal retrieval (image-to-music and music-to-image) and intra-modal retrieval (image-to-image and music-to-music) results
for the three different audio encoder models introduced in \autoref{subsec:Feature Extraction}. For the music-specific models, we include results for both frozen and unfrozen models, while we keep CLAP frozen throughout our experiments.

For image-to-music retrieval, the frozen Harmonic CNN and CLAP models perform the best. For music-to-image retrieval, the unfrozen Harmonic CNN and CLAP models perform the best. Interestingly, the unfrozen music-specific models generally perform better than their frozen counterparts for music-to-image retrieval, but this pattern does not hold for image-to-music retrieval. For music-to-music retrieval, CLAP consistently performs the best, demonstrating its ability to handle complex audio tasks while frozen.
Since Harmonic CNN and CLAP perform the best in cross-modal retrieval, we use these audio encoder models for the remainder of our analysis. We attribute the superior performance of Harmonic CNN and CLAP to their longer audio input lengths (5.0 seconds and 10.0 seconds) compared to Short-Chunk CNN's 3.7 seconds.

\begin{table}[t]
    \caption{Comparison with other emotion-matched cross-modal music retrieval works. We show results for our two best-performing image-to-music retrieval models. We report Precision@5 (P@5) and Mean Reciprocal Rank (MRR) metrics. All works use the AudioSet music mood subset \cite{Gemmeke-2017-AudioSet} for music retrieval.}
    \label{table:retrieval_comparison}
    \footnotesize
    \centering
    
    \begin{tabular}{l | l l | l l}
        \toprule
        
        Method & Input &
        Input Dataset &
        P@5 & MRR
        \\
        \midrule
        
        \multirow{2}{*}{\cite{Won-2021-emotion_embedding_spaces} V-A Regression} & \multirow{4}{*}{Text} &
        Alm's & 61.00\% & 73.98\%
        \\
        & & ISEAR & 62.18\% & 70.75\%
        \\
        \multirow{2}{*}{\cite{Won-2021-emotion_embedding_spaces} Metric Learning} & &
        Alm's & 51.56\% & 58.80\%
        \\
        & & ISEAR & 60.19\% & 66.75\%
        \\
        \midrule
        
        \multirow{2}{*}{\cite{Doh-2023-speech_to_music} Triplet + EmoSim} & \multirow{2}{*}{Speech} &
        IEMOCAP & 68$ \pm $3\% & 76$ \pm $3\%
        \\
        & & RAVDESS & 67$ \pm $2\% & 75$ \pm $3\%
        \\
        \midrule
        
        Emo-CLIM (HCNN) & \multirow{2}{*}{Image} & \multirow{2}{*}{DeepEmotion} &
        65.94\% & \textbf{78.59\%}
        \\
        Emo-CLIM (CLAP) & & & \textbf{68.15\%} & 76.65\%
        \\
        
        \bottomrule
    
    \end{tabular}

\end{table}

\textbf{Comparison With Other Works}:
Since emotion-matched image-to-music retrieval has not been addressed before in the literature, there is no direct baseline for comparison. Hence, we compare with two other works on emotion-matched music retrieval from other modalities: text-to-music retrieval \cite{Won-2021-emotion_embedding_spaces} and speech-to-music retrieval \cite{Doh-2023-speech_to_music}. We compare our image-to-music retrieval results---using our two best-performing frozen audio encoder models (where \emph{HCNN} = Harmonic CNN)---with these two works in \autoref{table:retrieval_comparison}.

For the text-to-music retrieval paper \cite{Won-2021-emotion_embedding_spaces}, we include results for their manual emotion label mapping\footnote{The authors report results for 3 different emotion label taxonomy mappings: valence-arousal-based, Word2Vec-embedding-based, and manual.}, since we also use a manual mapping. We report results for two of their best-performing methods along with the text datasets used.
For speech-to-music retrieval \cite{Doh-2023-speech_to_music}, we show results for the IEMOCAP \cite{Busso-2008-IEMOCAP} and RAVDESS \cite{Livingstone-2018-RAVDESS} speech datasets. We do not include results for the Hi,KIA dataset \cite{Kim-2022-HiKIA} because it is challenging to provide statistically meaningful results with the limited number of samples in this dataset (only 488 utterances in total).
We report results for their best-performing non-fusion-based models to provide a fair comparison with the rest of the models (which are all non-fusion-based).

\autoref{table:retrieval_comparison} suggests that our image-to-music retrieval framework substantially outperforms the text-to-music retrieval work presented in \cite{Won-2021-emotion_embedding_spaces}.
Our results are comparable to those of the speech-to-music retrieval work \cite{Doh-2023-speech_to_music}. We argue that aligning speech and music is more trivial than aligning images and music, because speech and music belong to the same modality (audio).
We recognize that these comparisons are not direct---due to the differences in modalities, datasets,
and emotion label taxonomies---but they provide useful insights into the effectiveness of our approach.

\subsection{Automatic Music Tagging}
\label{subsec:Automatic Music Tagging}

\textbf{Experimental Setup}:
Following previous studies \cite{Spijkervet-2021-CLMR, Avramidis-2023-VCMR}, we use automatic music tagging as a downstream task to evaluate our music representations. Music tagging is a multi-label classification task that aims to predict a number of semantic binary tags for a music track. These typically describe musical attributes such as genre, instrument, and mood. We use the popular MagnaTagATune dataset \cite{Law-2009-MagnaTagATune}, which consists of 25,000 music clips (around 30 seconds each) generated from 6,662 unique songs. In line with the literature \cite{Won-2020-music_tagging_evaluation, Spijkervet-2021-CLMR, Avramidis-2023-VCMR}, we select the top 50 most frequent tags for our evaluation.

To implement music tagging, we first generate music audio embeddings using the frozen pre-trained audio component of our Emo-CLIM framework. We then train a small classification head on these music embeddings using binary cross-entropy loss. The classification head consists of  a linear layer, a BatchNorm layer, a ReLU activation, and a second linear layer.
We evaluate performance using ROC-AUC and PR-AUC, averaged over all tags.

\textbf{Results}:
In \autoref{table:music_tagging} we present music tagging results on the MagnaTagATune dataset. We include two fully-supervised baselines: Short-Chunk CNN \cite{Won-2020-music_tagging_evaluation} and Harmonic CNN \cite{Won-2020-HarmonicCNN} for reference. In addition, we show the performance of three different models that are pre-trained on a self-supervised learning (SSL) task:
CLMR \cite{Spijkervet-2021-CLMR}, VCMR \cite{Avramidis-2023-VCMR}, and CLAP \cite{Wu-2023-CLAP}. We report results for CLMR and VCMR from their respective papers, but we implement this same approach for CLAP since it has not been done before. We include these SSL baselines since their training procedures
are similar to the one used in our approach. The three bottom rows show the results for Emo-CLIM, using three different audio encoder variants. Emo-CLIM performs on par with the SSL baselines, demonstrating that the emotion-aligned music embeddings are effective in capturing general music semantics in addition to affective information.

\begin{table}[t]
    \caption{Music tagging (MT) results on MagnaTagaATune \cite{Law-2009-MagnaTagATune}. First two rows show fully-supervised baselines, middle three rows show self-supervised (SSL) baselines, and the last three rows show the Emo-CLIM framework with three different audio encoder variants ($^{\dagger}$denotes unfrozen model). We report ROC-AUC and PR-AUC evaluation metrics, which are the standard in music tagging.}
    \label{table:music_tagging}
    \footnotesize
    \centering
    
    \begin{tabular}{l l | l l}
        \toprule
        
        Method & Pretraining Task & ROC-AUC & PR-AUC
        \\
        \midrule
        
        Short-Chunk CNN \cite{Won-2020-music_tagging_evaluation} & MT & 91.29\% & 46.14\%
        \\
        Harmonic CNN \cite{Won-2020-HarmonicCNN} & MT & 91.27\% & 46.11\%
        \\
        \midrule
        
        CLMR \cite{Spijkervet-2021-CLMR} & SSL & 89.3\% & 36.0\%
        \\
        VCMR \cite{Avramidis-2023-VCMR} & SSL & 89.08\% & 35.27\%
        \\
        CLAP \cite{Wu-2023-CLAP} & SSL & \textbf{91.04\%} & \textbf{39.40\%}
        \\
        \midrule
        
        Emo-CLIM (HCNN) & MT $ \rightarrow $ SupCon & 89.70\% & 36.00\%
        \\
        Emo-CLIM (HCNN$^{\dagger}$) & MT $ \rightarrow $ SupCon & 88.55\% & 33.80\%
        \\
        Emo-CLIM (CLAP) & SSL $ \rightarrow $ SupCon & 89.94\% & 37.12\%
        \\
        
        \bottomrule
    
    \end{tabular}

\end{table}

\section{Conclusion}
\label{sec:Conclusion}

In this work, we presented Emo-CLIM, a framework for learning an affective alignment between images and music audio. By using our proposed emotion-supervised contrastive loss, Emo-CLIM successfully learns an emotion-aligned image-music embedding space. We demonstrated that this joint embedding space is effective for cross-modal and intra-modal retrieval tasks, where the goal is to retrieve emotionally-relevant images or music clips. Furthermore, the learned music embeddings effectively capture general music semantics, as shown in the automatic music tagging evaluation.

In the future, we will incorporate emotion class similarities into our contrastive loss in order to improve the aligned representations. In addition, we plan to explore the effect of adding data augmentations to our training pipeline.
Our approach showed promising results for cross-modal affective alignment, and we hope that our work can help motivate further research in this exciting area.


\clearpage

\bibliographystyle{IEEEbib}
\bibliography{references}

\end{document}